# The Linear Relationship between Temporal Persistence, Number of Independent Features and Target EER

Lee Friedman, Hal S. Stern and Oleg V. Komogortsev, *Member, IEEE*

*Abstract*—If you have a target level of biometric performance (e.g. EER = 5% or 0.1%), how many units of unique information (uncorrelated features) are needed to achieve that target? We show, for normally distributed features, that the answer to that question depends on the temporal persistence of the feature set. We address these questions with synthetic features introduced in a prior report. We measure temporal persistence with an intraclass correlation coefficient (ICC). For 5 separate EER targets (5.0%, 2.0%, 1.0%, 0.5% and 0.1%) we provide linear relationships between the temporal persistence of the feature set and the log10(number of features). These linear relationships will help those in the planning stage, prior to setting up a new biometric system, determine the required temporal persistence and number of independent features needed to achieve certain EER targets.

## I. INTRODUCTION

WHEN planning a new biometric system, there are at least two key questions: (1) What is my target level of biometric performance? (2) How many features will I need to achieve a particular level of biometric performance? We define biometric performance in terms of equal error rate although other targets can be employed. It makes intuitive sense that sets of features which are less correlated, and therefore provide more additional information, will be most helpful. It also makes intuitive sense that more permanent features will be more useful than less permanent features. In this paper, we use normally distributed synthetic data to formally, explicitly and mathematically explore these relationships. We define unique features as features that are not intercorrelated, and we quantify permanence (we prefer the term "temporal persistence") with the intraclass correlation coefficient (ICC, [1]). The ICC ranges from 0 to 1.0, with 1.0 indicating perfect permanence (not change from time 1 to time 2). We have uncovered simple linear equations that relate the number of features (log10) to the temporal persistence of the features given a particular EER target. We are not aware of any prior reports that have addressed this specific question.

In a prior report we have introduced the concept of the ICC as an index of temporal persistence [1] for normally distributed features. We showed, for 12 of 14 different datasets, from several different biometric modalities, that choosing only the most temporally persistent features yielded the best biometric performance. In that paper we also showed that the improvement in performance with more temporally persistent features occurred because the genuine similarity score distributions for such features had higher medians and smaller inter-quartile ranges (IQR). This relationship was more clearly demonstrated in a subsequent theoretical paper (Friedman et al, submitted).

## II. METHODS

### A. Creation of Synthetic Features

In a prior report (Friedman et al., submitted) we fully describe our procedure to create synthetic features and we also included r code [2] to create such synthetic features. Briefly, let us assume that we want to model 10,000 subjects with 10 features and two sessions (occasions). First, we fill session 1 features with random draws from a normal distribution with a mean of 0 and an SD = 1. Next, we set session 2 data equal to session 1 data. At this point the ICC = 1.0, i.e., perfect agreement. Next, we add noise to both sessions. If your target ICC is known (e.g., 0.7), the SD of the added noise is:

$$\sqrt{(1 - ICC_{Target})/ICC_{Target}} \qquad (1)$$

$$\sqrt{\frac{0.3}{0.7}} = 0.65 \qquad (2)$$

In this way we can create synthetic datasets with features that

Date of first submission: 9/15/2018. The study was funded by 3 grants to Dr. Komogortsev: (1) National Science Foundation, CNS-1250718, www.NSF.gov; (2) National Institute of Standards and Technology, 60NANB15D325, www.NIST.gov; (3) National institute of Standards and Technology, 60NANB16D293. Dr. Stern's contributions were partially funded through Cooperative Agreement # 70NANB15H176 between NIST and Iowa State University, which includes activities carried out at Carnegie Mellon University, University of California Irvine, and University of Virginia.

L. Friedman is with the Department of Computer Science at Texas State University, San Marcos, TX, USA, 78666 (e-mail: lfriedman10@gmail.com).

H. S. Stern is Chancellor's Professor, Department of Statistics, Donald Bren School of Information and Computer Sciences, University of California, Irvine, CA, USA, 92697 (e-mail: sternh@uci.edu).

O.V. Komogortsev is with the Department of Computer Science and Engineering, Michigan State University, East Lansing, MI, USA, 48824 (e-mail: ok@msu.edu).

are uncorrelated (except to a very small extent due to chance), with as many subjects and features as we like, with a particular level of temporal persistence.

### B. Creation of Sets of Features

We generate a series of synthetic data sets with varying ICCs. Specifically, we created 7 different datasets with each dataset consisting of 350 features, 10,000 subjects, and 2 sessions. The features in each dataset were created with 7 different ICC target values (0.35, 0.45, 0.55, 0.65, 0.75, 0.85, 0.95), but due to chance, there is some variability around these ICC targets (Fig. 1), although their mean ICCs are very close to the target (Table I).

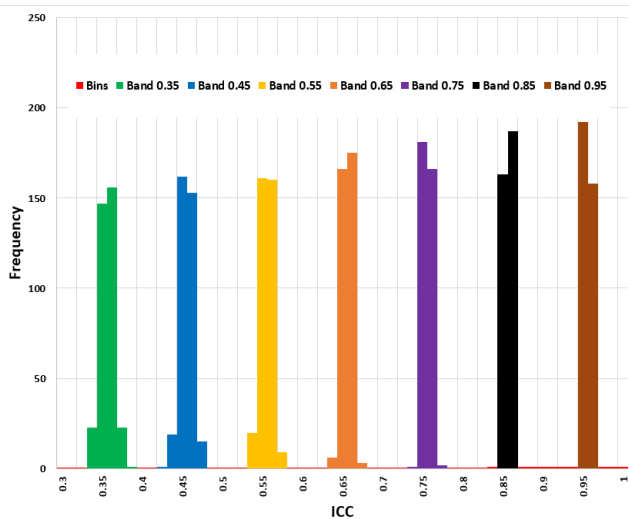

*Figure 1: Frequency histograms of ICC in each Band.*

| Table I Mean (SD) ICC for Each Band | | |
|---|---|---|
| Band | ICC Mean | ICC SD |
| 0.35 | 0.350 | 0.009 |
| 0.45 | 0.450 | 0.008 |
| 0.55 | 0.549 | 0.007 |
| 0.65 | 0.650 | 0.006 |
| 0.75 | 0.750 | 0.004 |
| 0.85 | 0.850 | 0.003 |
| 0.95 | 0.950 | 0.001 |

### C. Procedure

The goal was to determine how many features in each Band were required to achieve one of 5 target EER values (5.0%, 2.0%, 1.0%, 0.5% and 0.1%). Initially, we computed EERs for each band for from 1 to 350 features, with features chosen randomly from the set of 350. We did this initially for a single repetition. Based on the results, we were able to find certain ranges of features numbers that were approximately near where each EER target was crossed. On the next iteration, we did a more limited search with 20 replications. Based on these results, we further limited the search but now performed 100 iterations.

### III. RESULTS

With 100 iterations, it was possible to definitively determine the number of features required to cross below the 5 EER target levels. These results are presented in Table II.

| Table II. Number of Features Required to Achieve Particular EER Targets for Each Band | | | | | |
|---|---|---|---|---|---|
| Band | EER < 5.0 | EER < 2.0 | EER < 1.0 | EER < 0.50 | EER < 0.10 |
| 0.35 | 82 | 127 | 162 | 198 | 281 |
| 0.45 | 48 | 74 | 94 | 115 | 166 |
| 0.55 | 30 | 46 | 59 | 72 | 102 |
| 0.65 | 19 | 30 | 38 | 46 | 66 |
| 0.75 | 13 | 20 | 25 | 30 | 43 |
| 0.85 | 8 | 12 | 16 | 19 | 27 |
| 0.95 | 5 | 7 | 8 | 10 | 14 |

When we plotted the number of features in Table II against the ICC Band targets, the relationships were complex. However, when we took the log10 of the number of features we noted that now the data looked like a series of 5 parallel lines (Fig. 2).

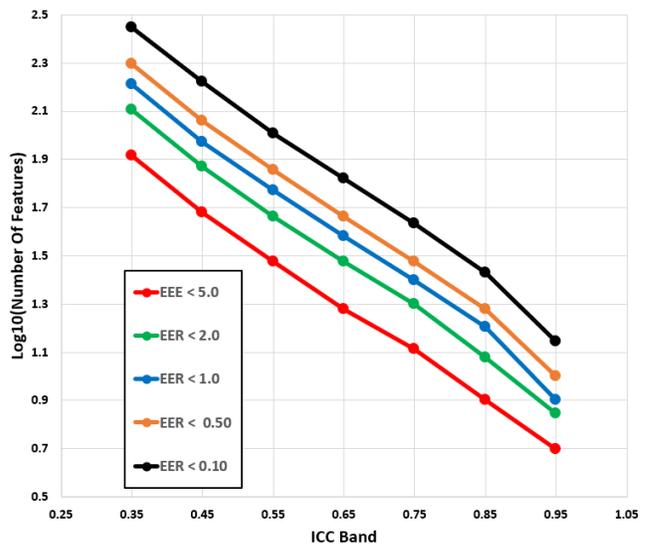

*Figure 2: Linear relationships between ICC and log10(number of features) for 5 target EER levels. When interpreting this graph, it may be useful to consider that, as the ICC of the features increase, the EER step from N Features to N+1 Features is larger, and so there is less "EER resolution" with higher ICC bands.*

The regression results are displayed in Table III. The F-values and R-Sqr values were extremely large and the p-values were all extremely small. As evident in Fig. 2, the slopes are very similar, especially for EER targets from 2.0% to 0.1%, and the lines differ mostly in terms of the intercept.

We also looked at the same relationships using different error thresholds. If we define the False Reject Rate (FRR) as the percentage of genuine similarity scores at or below a given and the False Acceptance Rate (FAR) as the number of impostor scores above a given thresholds, we determined the number of features required to obtain a FRR = 1% at the following FAR levels (0.1%, 0.01%, 0.001% and 0.0001%). Our results are presented in Figure 3 and Table 4.

| Table III: Regression Results | | | | | | |
|---|---|---|---|---|---|---|
| EER Target | F | df | p-value | R-Sqr | Slope | Intercept |
| 5.00% | 3637 | 1 | 2.E-08 | 0.999 | -1.987 | 2.587 |
| 2.00% | 2801 | 1 | 5.E-08 | 0.998 | -2.042 | 2.804 |
| 1.00% | 1036 | 1 | 5.E-07 | 0.995 | -2.082 | 2.930 |
| 0.50% | 1678 | 1 | 2.E-07 | 0.997 | -2.084 | 3.016 |
| 0.10% | 1476 | 1 | 2.E-07 | 0.997 | -2.093 | 3.176 |

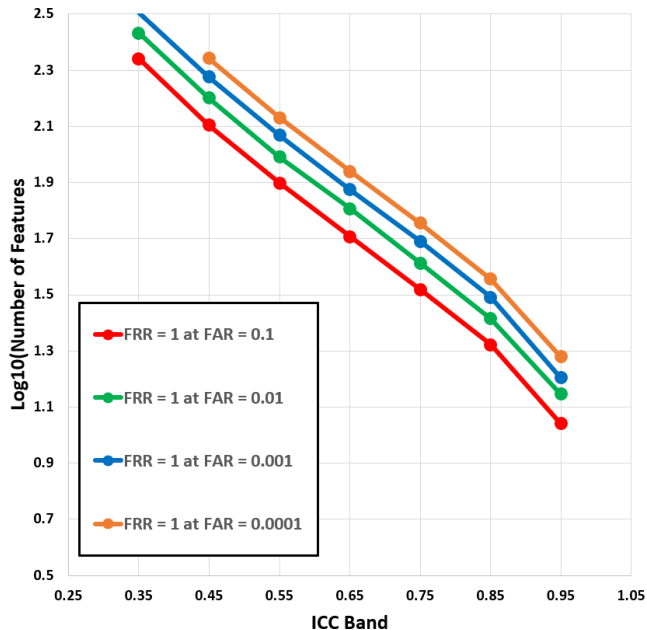

*Figure 3: Linear relationships between ICC and log10(number of features) for 4 target rates (see box in plot).*

| Table IV: Regression Results | | | | | | |
|---|---|---|---|---|---|---|
| Target | F | df | p-value | R-Sqr | Slope | Intercept |
| FRR = 1.0 at FAR = 0.1 | 1605 | 1 | 2.E-07 | 0.997 | -2.086 | 3.060 |
| FRR = 1.0 at FAR = 0.01 | 2089 | 1 | 9.E-08 | 0.998 | -2.076 | 3.150 |
| FRR = 1.0 at FAR = 0.001 | 1454 | 1 | 2.E-07 | 0.997 | -2.091 | 3.232 |
| FRR = 1.0 at FAR = 0.0001 | 871 | 1 | 8.E-06 | 0.995 | -2.064 | 3.279 |

### IV. Discussion

We describe linear relationships between the temporal persistence of a set of features and the number of independent features (log10) required to achieve 5 common EER targets (5.0%, 2.0%, 1.0%, 0.5% and 0.1%). As far as we can tell, no one has ever even framed a question in this manner, much less reported on similar relationships. In the first instance, we believe that our results will provide insight to biometric researchers about these relationships. Although real data features are not uncorrelated, a decorrelation step can be employed to decorrelate real features. (The decorrelation step uses an inverse Cholesky factorization [3]). Since assessing temporal persistence using the ICC is only recently introduced to the biometric community [1] few researchers are aware of the temporal persistence of their feature sets. We believe the present results provide further motivation for biometric researchers to begin to adopt this or other temporal persistence indices (especially for non-normally distributed, ordinal or nominal data types). Our results could provide insights to those in the planning stage of creating a new biometric system. If they can specify the target EER and the approximate temporal persistence of their features, our equations can tell them how many such independent features will be required.

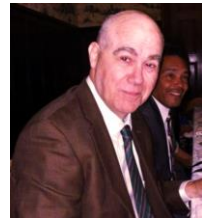

**Lee Friedman** Dr. Friedman has a B.A. in psychology from the University of Illinois, Champagne, Illinois, USA (1975), an M.A. in biological psychology from the University of Chicago, Chicago Illinois, USA (1979) and a Ph.D. in Neuroscience from McGill University, Montreal Canada (1983).

Most of his career he worked as a clinical neuroscientist studying the pathophysiology of major mental illness. Currently he is working in the area of oculomotor biometrics and eye movements. He has 74 peer-reviewed publications.

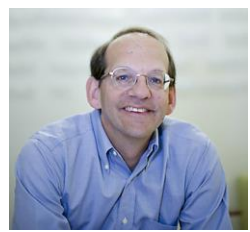

**Hal S. Stern** is Chancellor's Professor, Department of Statistics, Donald Bren School of Information and Computer Sciences, University of California, Irvine. He has a B.S. in mathematics from the Massachusetts Institute of Technology, Cambridge, MA, USA (1981), an M.S. (1985) and Ph.D. (1987) in statistics from Stanford University, Stanford, CA, USA. His research interests include Bayesian methods, model diagnostics, forensic statistics, and statistical applications in biology/health, social sciences, and sports. He is a Fellow of the American Association for the Advancement of Science, the American Statistical Association and the Institute of Mathematical Statistics.

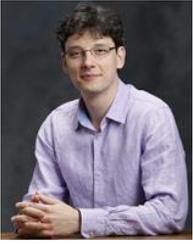

**Oleg V. Komogortsev**

Dr. Komogortsev is currently a tenured Associate Professor at Michigan State University. Dr. Komogortsev has received his B.S. in Applied Mathematics from Volgograd State University, Russia, and M.S./Ph.D. degree in Computer Science from Kent State University, Ohio. Dr. Komogortsev conducts research in eye tracking with a focus on cyber security (biometrics), human computer interaction, usability, bioengineering, and health assessment. This work has thus far yielded more than 100 publications and several patents. Dr. Komogortsev's research was covered by the national media including NBC News, Discovery, Yahoo, Livesience and others. Dr. Komogortsev is a recipient of two Google Virtual Reality Research Awards and a Google Faculty Research Award. Dr. Komogortsev has also won National Science Foundation CAREER award and Presidential Early Career Award for Scientists and Engineers (PECASE) on the topic of cybersecurity with the emphasis on eye movement-driven biometrics and health assessment. In addition, his research is supported by the National Institute of Standards, Sigma Xi the Scientific Research Society, and various industrial sources. Dr. Komogortsev's current grand vision is to push forward eye movement-driven biometrics as person recognition and health assessment platform in the future virtual and augmented reality devices.